\documentclass[aps,pra,amsmath,amssymb,showpacs,twocolumn,preprintnumbers,
floatfix,letterpaper]{revtex4-1}

\usepackage{bm}
\usepackage{epsfig}
\usepackage{graphicx}
\usepackage{graphics}
\usepackage{amsmath}
\usepackage{amssymb}
\usepackage{amsfonts}
\usepackage{color}

\newcommand{\be}{\begin{equation}}
\newcommand{\ee}{\end{equation}}
\newcommand{\ba}{\begin{eqnarray}}
\newcommand{\ea}{\end{eqnarray}}
\newcommand{\br}{\begin{array}}
\newcommand{\er}{\end{array}}
\newcommand{\Schro}{Schr\"o\-din\-ger }

\def\rgb#1#2#3{}

\def\black{\rgb{0}{0}{0}}

\begin{document}
\preprint{Draft/RevTeX 4/Xiaoxu, Klaus, Barry, 17~July~2010}
\author{Xiaoxu~Guan$^1$}
\author{Klaus~Bartschat$^1$}
\author{Barry I. Schneider$^2$}
\affiliation{$^1$Department of Physics and Astronomy, Drake University, Des Moines, Iowa 50311, USA}
\affiliation{$^2$Physics Division, National Science Foundation, Arlington, Virgina 22230, USA}

\date{\today}

\title{\bf Two-photon Double Ionization of H$_2$ in Intense Femto\-second Laser Pulses}

\begin{abstract}
Triple-differential cross sections for two-photon double ionization of molecular hydrogen are presented 
for a central photon energy of $30$~eV. The calculations are based on a fully {\it ab initio}, non\-perturbative, 
approach to the time-dependent \Schro equation in prolate spheroidal coordinates, discretized by a finite-element 
discrete-variable-representation.  The wave function is propagated in time for a few femtoseconds using the short,
iterative Lanczos method to study the correlated response of the two photo\-electrons to short, intense laser radiation.  
The current results often lie in between those of Colgan {\it et al} [J. Phys. B
{\bf 41} (2008) 121002] and Morales {\it et al} [J. Phys. B {\bf 41} (2009) 134013]. 
However, we argue that these individual predictions should not be compared directly to each other, 
but preferably to experimental data generated under well-defined conditions.
\black

\end{abstract}

\maketitle

Mapping the four-body breakup processes of the hydrogen 
molecule via multiphoton absorption has recently become feasible experimentally
at the free-electron laser facility FLASH.
Although the angular distributions and the kinetic-release-energy spectrum of
ionic fragments of D$_2$ were measured~\cite{Jiang2010-1}, 
determining the angular distribution of the ejected electrons is very difficult.   
Compared to one-photon double ionization (DI) of H$_2$~\cite{Vanroose2006,Colgan2007}, an accurate theoretical description of the 
two-photon DI of the H$_2$ molecule is also extremely challenging.   Previous calculations by Colgan~{\it et al}~\cite{Colgan2008}, who employed the 
time-dependent close-coupling (TDCC) method, and by Morales~{\it et al}~\cite{Morales2009}, who used a 
time-independent exterior complex scaling (ECS) treatment, showed considerable disagreements regarding both the shape and the magnitude 
of the predicted triple differential cross section (TDCS). However, both predicted a strong dependence of the 
angular distribution of the ejected electrons on the alignment angle between 
the linear laser polarization vector $(\bm{\epsilon})$ and the molecular axis
$(\bm{\varsigma})$.

The discrepancies between the previous, computationally very demanding
calculations provided the primary motivation for the present project. We emphasize, however,
that these calculations were performed for much different
conditions.  Colgan~{\it et al}~\cite{Colgan2008}
used a femto\-second laser pulse with a peak laser intensity of 10$^{15}\,$W/cm$^2$ that could lead to serious
depletion of the ground state and {\it may} not be in the perturbative regime.  The
calculation of Morales {\it et al}~\cite{Morales2009} was for a well-defined
photon energy and is equivalent to a weak pulse of near infinite duration.  
In light of the ongoing discussions, it seemed important to perform another
independent calculation to
investigate the similarities and differences in the results and to 
understand their origin. Consequently,
we also consider the two-photon DI of the H$_2$ molecule in the parallel and
perpendicular geometries at an incident photon energy centered around $30$~eV and 
equal sharing of the excess energy.
Our pulse (see details below) lies in the perturbative regime but, like the TDCC model, has a bandwidth that may
access doubly excited electronic states in the H$_2$ spectrum unavailable to Morales {\it et al}.
The effect of such states on the outcome of a short-pulse experiment is not known at this time.  The details could be highly complex, and 
a theoretical treatment may ultimately have to go beyond the current model.

The two-center nature of H$_2$ already destroys the spherical symmetry of the problem even in the absence
of an external laser field.  While it is numerically feasible to carry out the calculations in a spherical coordinate
system at the center of the H$_2$ molecule, this becomes more and more difficult for heavier diatomics due to the slow
convergence of the multi-center electron-nuclear interaction.  Fortunately, the prolate spheroidal coordinate system
offers an attractive alternative for diatomics~\cite{Tao2009}, since the electron-nuclear interaction is rendered benign.

The addition of an external field causes a symmetry reduction when the laser field is not parallel to the internuclear axis.  
Compared to the atomic He target, therefore, the situation in this simplest two-electron molecule is complicated due to an additional degree of freedom, 
namely the alignment angle between the $\bm{\epsilon}$- and
$\bm{\varsigma}$-axes. While the emission
modes of photo\-electrons from atomic targets exhibit rotational symmetry
with respect to the polarization vector, this may no longer be the case for 
an arbitrary $\bm{\epsilon}-\bm{\varsigma}$ geometry.
All of these issues conspire to make the
response of the H$_2$ molecule to temporal laser fields much more computationally demanding than the two-photon DI of the helium
atom. 
 
For the frozen-nuclei approximation used here and in~\cite{Colgan2008,Morales2009} to be physically meaningful, 
the time scale for ionization by the field has to be much shorter than the characteristic time of the
nuclear motion. For {\it direct}, rather than {\it sequential\/} double ionization, this condition is fulfilled here, since the
{\it simultaneously\/} ejected electrons have a speed over ten times larger 
than what the protons can achieve in the subsequent Coulomb explosion.
For a central photon energy of $30$~eV, we expose
the H$_2$ molecule to a \hbox{$10$-cycle} sine-squared laser pulse at a peak intensity of $10^{14}\,$W/cm$^2$, followed by a \hbox{$2$-cycle} field-free
evolution. This ``$10+2$''-cycle takes
about $1.6$~fs, which 
is sufficiently long to extract a well-defined
cross section in the perturbative regime of relatively low peak intensities.
Nevertheless, a comparison with results from time-independent calculations is {\it not} directly possible, due to the energy width of several~eV.

Our treatment of the six-dimensional time-dependent \Schro equation (TDSE) of the laser-driven 
H$_2$ molecule relies on the {\em two-center} prolate spheroidal coordinates $(\xi,\eta,\varphi)$.
The ranges of the variables are $\xi\in[1,+\infty)$, $\eta\in[-1,+1]$, and $\varphi\in[0,2\pi]$.
For the temporal laser field ${\bm E}(t)$ = ${\bm\epsilon}E(t)$, 
the TDSE in the dipole length gauge reads 
\begin{equation}
i\frac{\partial}{\partial t}\Psi(1,2,t)=\Big[{\cal H}_1+{\cal H}_2
+\frac{1}{r_{12}} +\bm{E}(t) \cdot(\bm{r}_1+\bm{r}_2) \Big]\Psi(1,2,t).
\end{equation}
Here ${\cal H}_i$ represents the field-free single-electron Hamiltonian for electron~``$i$''
while ${\bm r}_i$ is its coordinate,
measured relative to the center of the molecule. As usual, $r_{12} = |\bm{r}_1- \bm{r}_2|$.

We expand the H$_2$ wave function in the body-frame as
\begin{equation}
\Psi(1,2,t) =\frac{1}{2\pi} \sum_{m_1m_2}\Pi_{m_1m_2}(\xi_1,\eta_1,\xi_2,\eta_2,t) 
         e^{ i ( m_1\varphi_1 + m_2 \varphi_2 )}, 
\end{equation}
where $m_1$ and $m_2$ denote the magnetic quantum numbers of the two electrons along 
the molecular axis.
To discretize this partial differential equation, we employ the finite-element discrete-variable-representation
(FEDVR) approach for the ($\xi$,$\eta$) variables~\cite{Tao2009}.  Thus  $\Pi_{m_1m_2}(\xi_1,\eta_1,\xi_2,\eta_2,t)$ is 
expanded in a product of ``radial" $\{f_{i}(\xi)\}$ and ``angular" $\{g_{k}(\eta)\}$ DVR
functions.  The notation originates from the asymptotic behavior of $\xi\rightarrow 2r/R$ and 
$\eta\rightarrow \cos\theta$ in spherical coordinates, when the electron is far
away from the nuclei.
 
To treat the electron-electron interaction we employ the Neumann expansion of
$1/r_{12}$ in prolate
spheroidal coordinates \cite{Morse1953},
\begin{align}
\label{ele}
\frac{1}{r_{12}}=&
\frac{2}{R}\sum_{\ell=0}^{\infty}\sum_{m=-\ell}^{\ell}(-1)^{|m|}(2\ell+1)
\bigg(\frac{(\ell-|m|)!}{(\ell+|m|)!} \bigg)^2 \\ \notag
\times& 
P_{\ell}^{|m|}(\xi_<)Q_{\ell}^{|m|}(\xi_>)P_{\ell}^{|m|}(\eta_1)P_{\ell}^{|m|}
(\eta_2)
e^{im(\varphi_1-\varphi_2)}.
\end{align}
Here $\xi_>$ and $\xi_<$ are the larger and smaller of $\xi_1$ and $\xi_2$,
respectively. $P_{\ell}^{|m|}(\xi)$ and $Q_{\ell}^{|m|}(\xi)$ are regular and
irregular Legendre functions defined for $\xi\in(1,+\infty)$, while
$P_{\ell}^{|m|}(\eta)$ is specified for $\eta\in[-1,+1]$. 
By suitably generalizing the approach used in~\cite{McCurdy2004} for the spherical case,
it is possible to reformulate the computation of the required matrix elements of
the operator in~Eq.~(\ref{ele})
as the solution of a two-center Poisson equation.  This results in a diagonal representation of the 
matrix elements of the electron-electron Coulomb interaction and considerably simplifies the
FE-DVR discretization procedure.  

The time-dependent electron wave packets of the laser-driven H$_2$ molecule were generated
by using our recently developed Arnoldi-Lanczos algorithm on the DVR mesh points~\cite{Guan2008}.
Compared to  the treatment of atomic targets in spherical coordinates, however,
the boundary conditions in spheroidal coordinates require further  
elaboration. As demonstrated in~\cite{Tao2009}, the ``radial'' and ``angular'' functions are not
simple polynomials for odd $|m|$.  Special attention must be paid to the
square-root
asymptotics as $\xi \rightarrow 1$ (from above) and
$|\eta|\rightarrow 1$ (from below) when $|m|$ is odd. By introducing the factors
$\sqrt{\xi^2-1}$ and $\sqrt{1-\eta^2}$, respectively, in the DVR bases $f_i(\xi)$
and $g_k(\eta)$ for odd $|m|$ to account for the behavior near the
boundary, it is possible to use a single quadrature for all~$m$.  

A spatial box with $\xi_{\rm max}=100$ was set up to truncate
the configuration space.
Solving the TDSE in imaginary time with a grid of $N_{\xi}\times N_{\eta}=224\times 9$ and
$|m|_{\rm max}=|m_1|_{\rm max}=|m_2|_{\rm max}=3$,
we obtained the electronic energy of the
initial $X\,^1\Sigma_g^+$ state at the internuclear distance of $R=1.4\, a_0$
($a_0 = 0.529 \times 10^{-10}\,$m denotes the Bohr radius)
as $-1.8873$ atomic units (a.u.), in good agreement with the recent benchmark
value of
$-1.888761428$ a.u.~by Sims and Hagstrom \cite{Sims2006} (after taking out the
 nucleus-nucleus interaction of~$1/1.4$). This yields a
double-ionization threshold of $51.4$~eV 
above the initial electronic state. 

To calculate the angle-resolved
differential cross sections,
we project the wave packet
at the end of the time propagation onto uncorrelated two-electron continuum states,
which are approximately constructed in terms of the single-electron continuum
states of the H$^+_2$ ion. For a momentum $k$, the ``radial'' part of the
continuum state, $T_{|m|q}^{(k)}(\xi)$, behaves asymptotically as
\begin{equation}
T_{|m|q}^{(k)}(\xi) \stackrel{\xi\rightarrow\infty}{\longrightarrow}
\frac{1}{\xi R}\sqrt{\frac{8}{\pi}}\sin\Big(c\xi+\frac{a}{2c}\ln(2c\xi)-\frac{\ell\pi}{2}
+\Delta_{|m|q}(k)\Big).
\end{equation}
Here $\Delta_{|m|q}(k)$ is the two-center Coulomb phase shift, \hbox{$c=kR/2$},
\hbox{$a=2R$}, and \hbox{$\ell=|m|+q$} for the H$_2^+$ ion. 
In addition to $|m|$, an integer $q$, which is the number of nodes of the 
angular function, is used to label the continuum state.
In the unified-atom limit,
\hbox{$R \rightarrow 0$}, the above two-center radial function reduces to the well-known
atomic Coulomb wave function.

Let us 
comment again on the similarities and the differences between the various approaches.  
One might expect our calculation to yield similar results to the TDCC model,
although our numerical implementation is very different from that of Colgan {\it et al}~\cite{Colgan2008}. For example,
they used a finite-difference method and a flat-top pulse with $12$-cycle time duration at a much stronger
intensity of $10^{15}\,$W/cm$^2$. 
Most importantly, however, both models project onto uncorrelated Coulomb
functions.

In the ECS approach of 
Morales {\it et al} \cite{Morales2009}, on the other hand, the information was
extracted from a time-independent wave function,
which effectively corresponds to an infinite propagation time.  The complicated
three-body correlated Coulomb boundary condition 
is avoided in this approach.   
We would best simulate the ECS model by exposing the H$_2$
molecule to a weaker light field with a longer pulse duration, and also by 
propagating for a long time after the external field
has died off. Then the electrons have moved very far away from the nuclei and the projection to uncorrelated 
functions is increasingly appropriate~\cite{Madsen2007}. Since this would require 
a large amount of computational resources, we decided on the compromise of testing our procedure by varying the number
of field-free propagation cycles between one and three. We only noticed a small sensitivity to the time of projection,
and hence are confident that our results are converged to a few percent.  In addition, it should be noted that TDCC and ECS results for
one-photon DI~\cite{Vanroose2006,Colgan2007}, performed with similar philosophies of extracting the information, showed excellent
agreement with each other.

\begin{figure}[tbh]
\centering
\epsfig{file=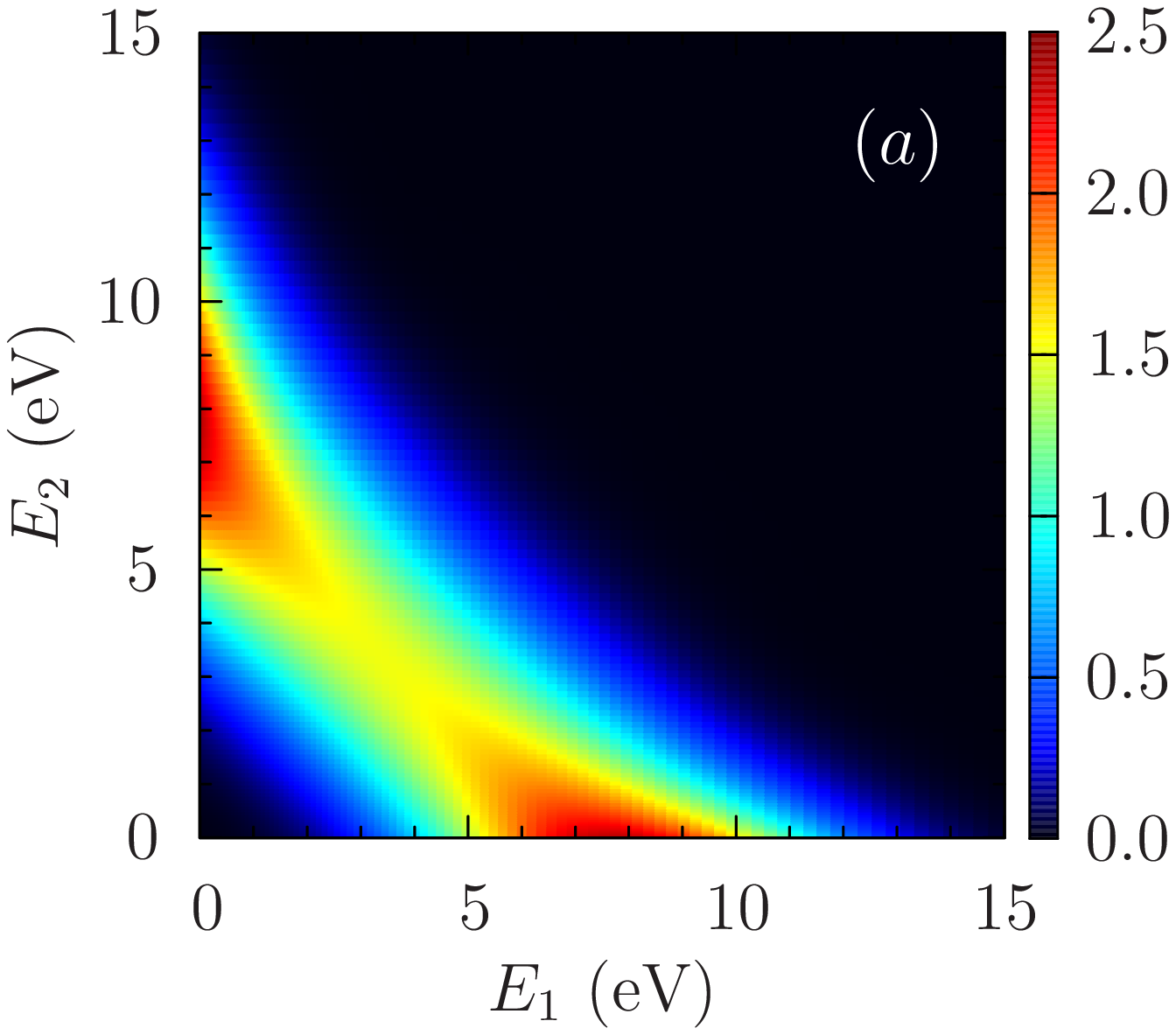,width=4.26cm,clip=}\hspace{0.2mm}
\epsfig{file=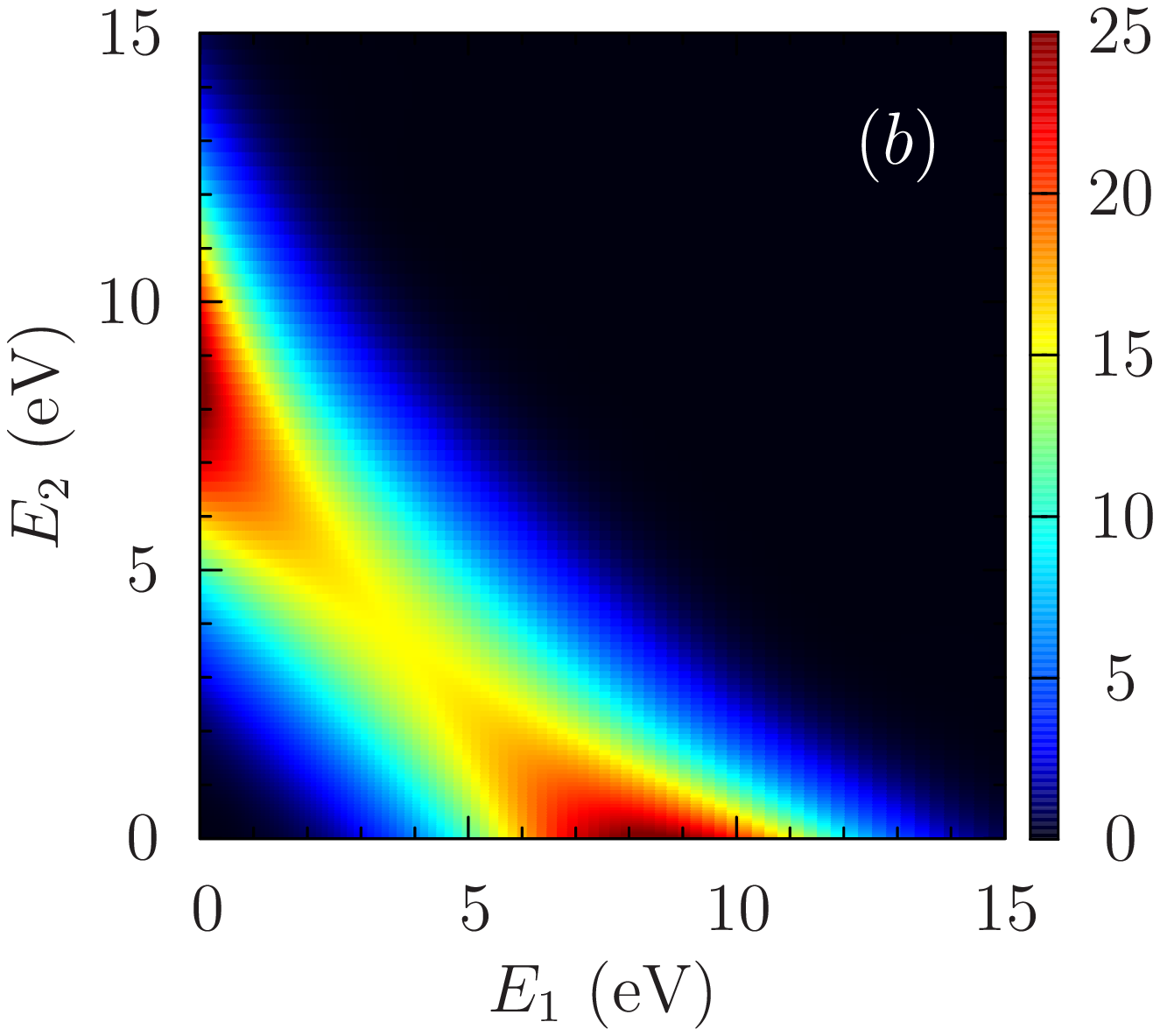,width=4.26cm,clip=}
\caption{(Color online) Energy probability distribution  of two ejected
electrons in the
parallel~(a) and perpendicular (b)~geometries for a
sine-squared laser pulse of $10$ optical cycles, a central energy of $30$~eV, and a peak intensity of
$10^{14}\,$W/cm$^2$.
The color bars correspond to multiples of $10^{-7}$ eV$^{-2}$.  Note the
different scales in the two panels.}
\label{fig:energy}
\end{figure}

Figure \ref{fig:energy} depicts the correlated energy distributions of the two
ejected
electrons after two-photon absorption. As they are
similar in shape for the both parallel and the perpendicular geometries,
information about any possibly preferred direction of the outgoing electrons 
is smeared out in the energy distributions.
Figure~\ref{fig:energy} reveals, however, that the probability of two-photon DI in the perpendicular geometry is about ten times larger
than for the parallel case, in qualitative agreement with the findings
of~\cite{Colgan2008}.

Next we explore the angle-resolved cross sections in
a particular plane, namely that formed by the molecular axis and the laser
polarization axis, a coplanar configuration. Note that the angles of
the two ejected electrons, $\theta_1$ and $\theta_2$, are measured with respect to
the $\bm{\epsilon}$ vector rather than the $\bm{\varsigma}$ axis. 

\begin{figure}[htb]
\centering
\epsfig{file=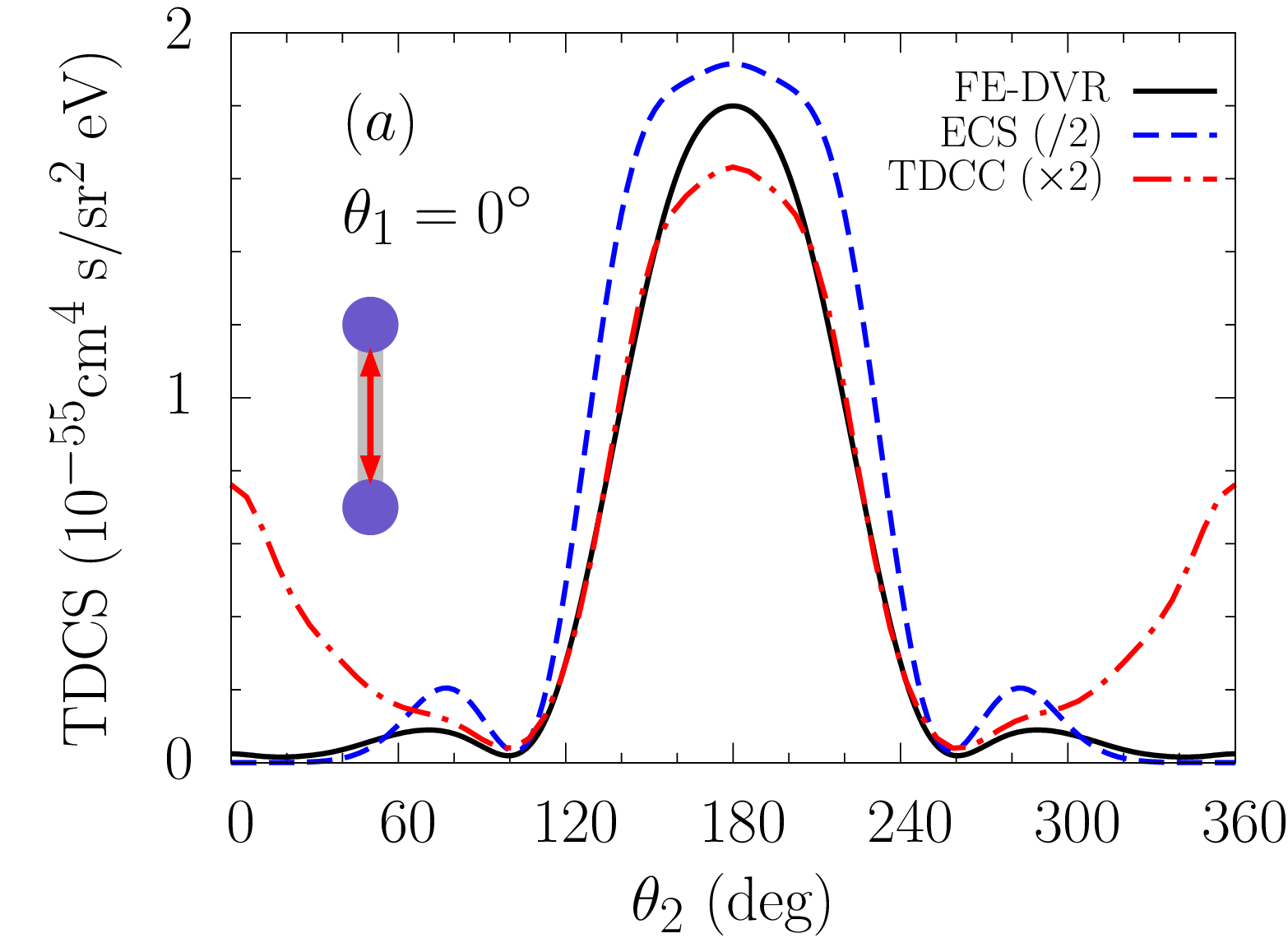,width=4.27cm,clip=}
\epsfig{file=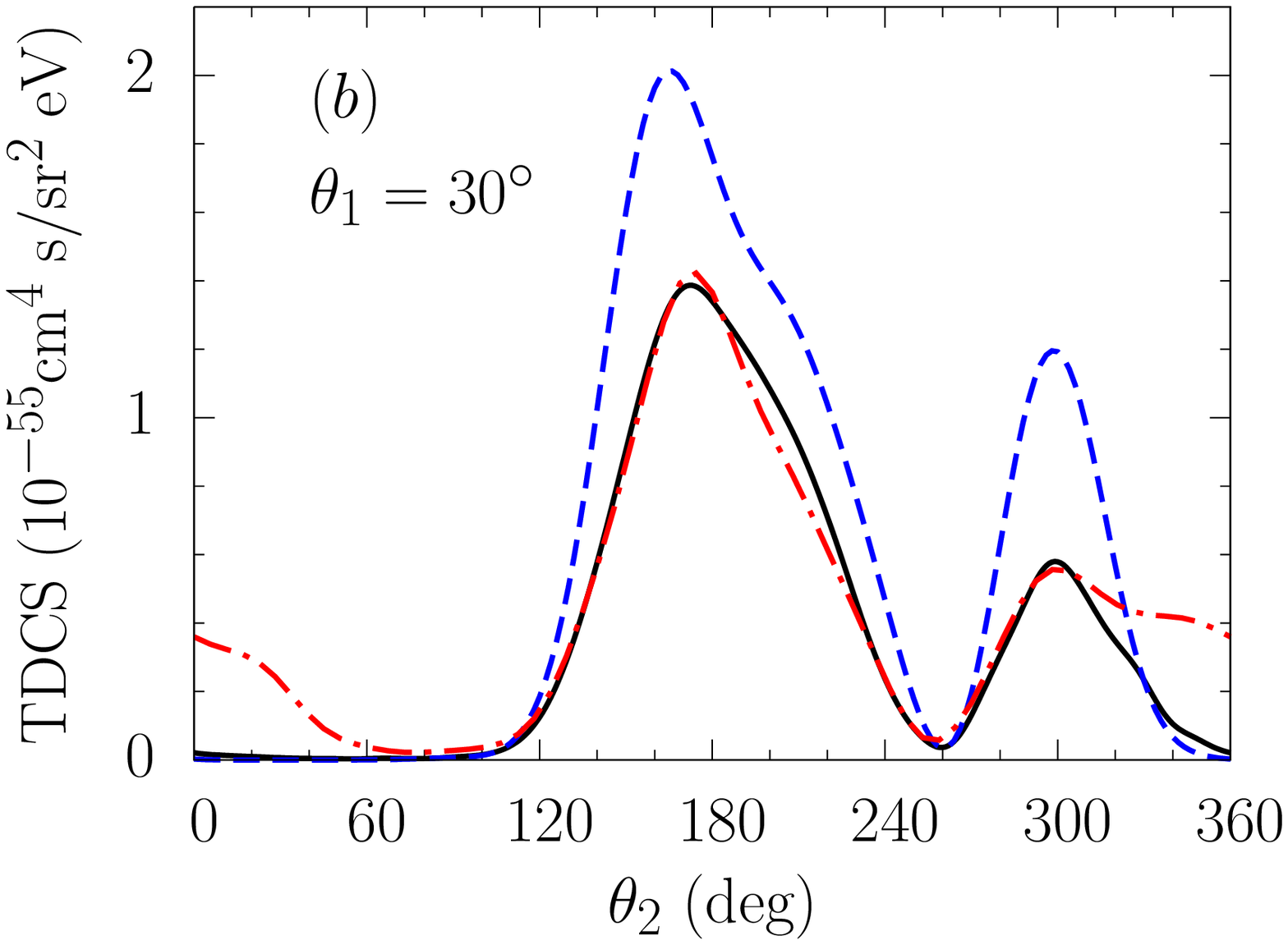,width=4.27cm,clip=} \\
\epsfig{file=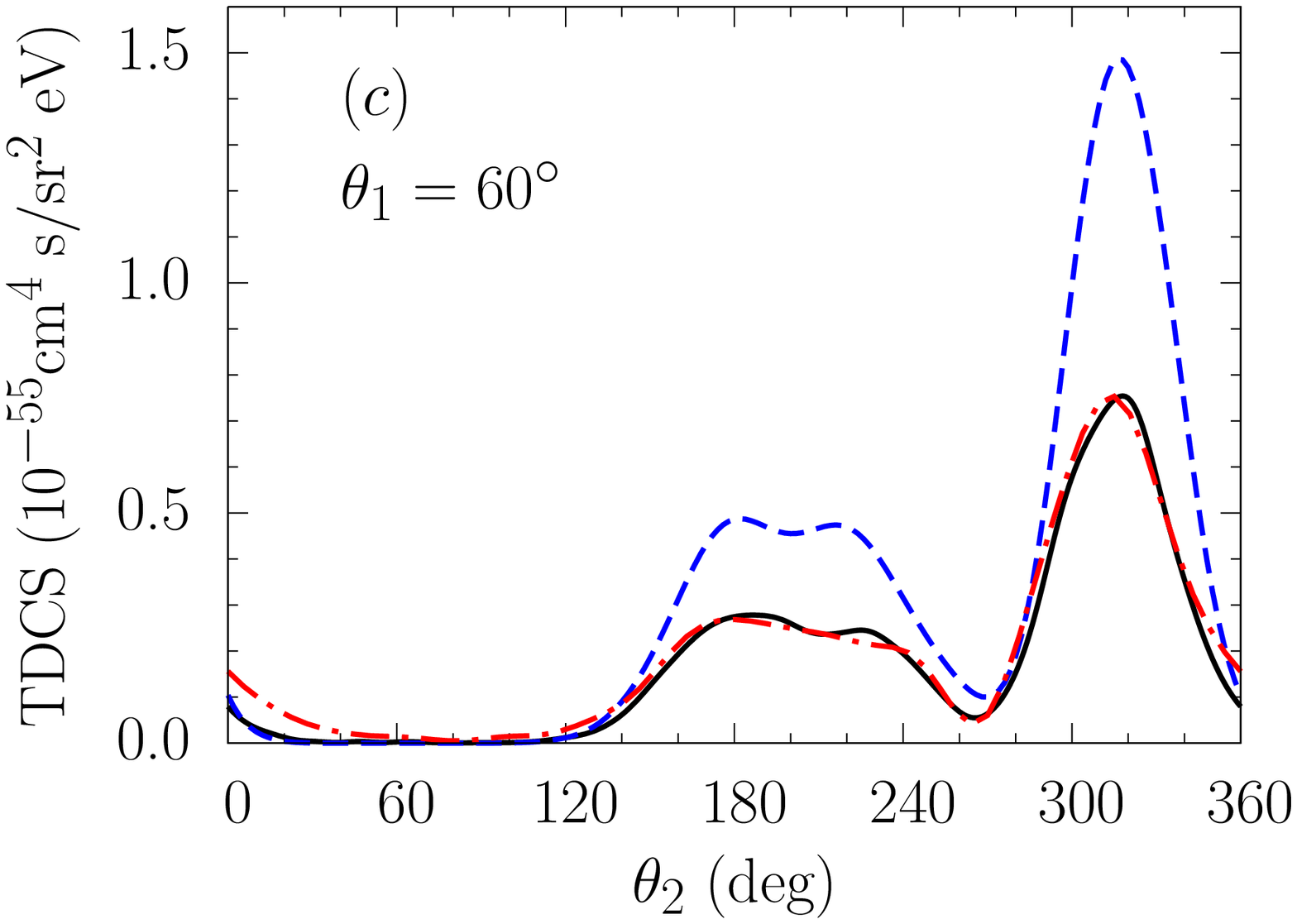,width=4.27cm,clip=}
\epsfig{file=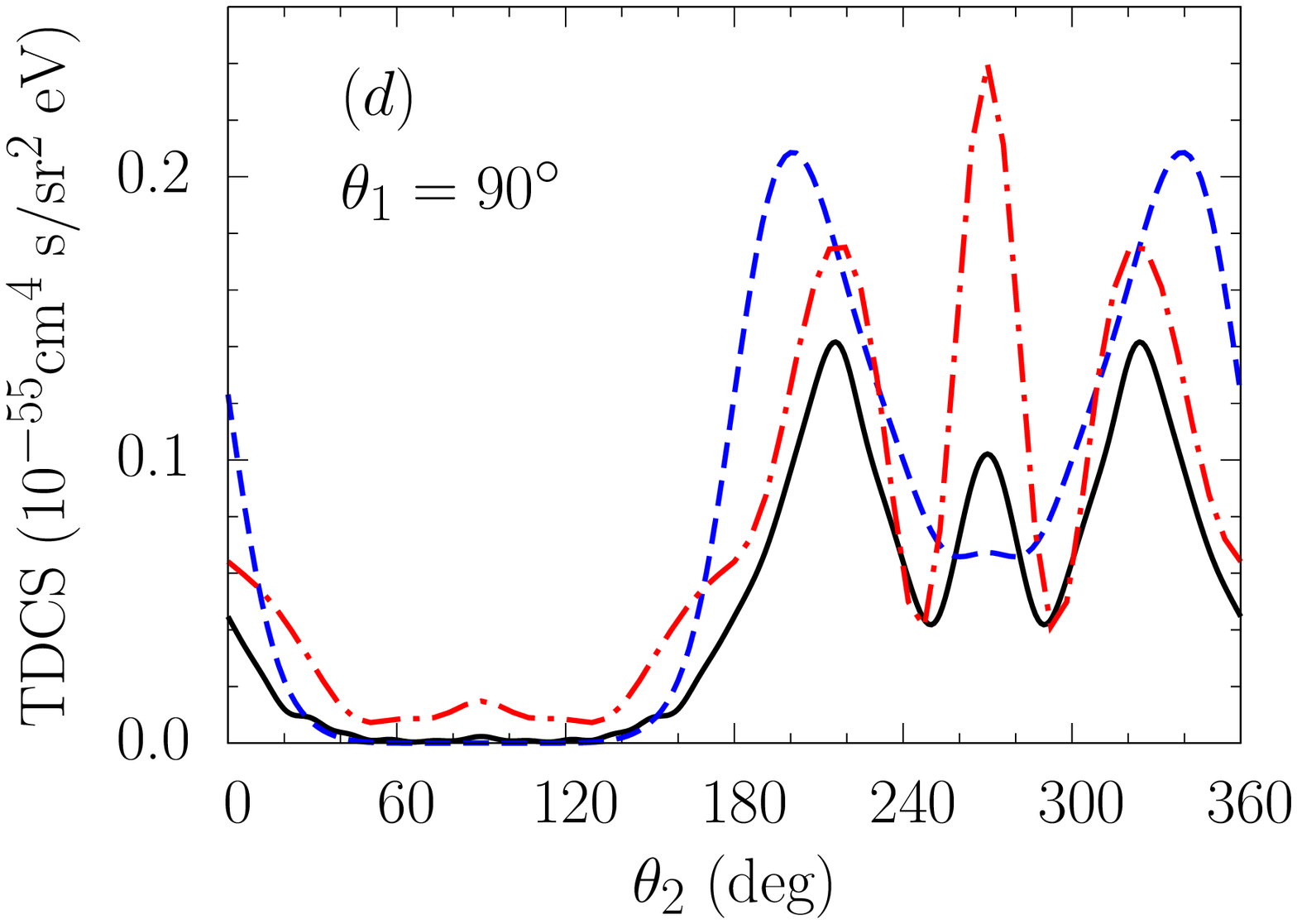,width=4.27cm,clip=}
\caption{(Color online) Coplanar TDCS for
two-photon DI of H$_2$ at equal energy sharing ($E_1=E_2=4.3$~eV) of the two
ejected electrons in the parallel geometry. The
laser
parameters are the 
same as in Fig.~\ref{fig:energy}. 
Also shown are the TDCC results of Colgan {\it et al} \cite{Colgan2008} and the
ECS
results of Morales {\it et al} \cite{Morales2009}, multiplied by the scaling
factors indicated in the legend. The definition of the TDCS is consistent in all three calculations.} 
\label{fig:tdcs-para}
\end{figure}

Our TDCS results for the parallel
geometry are shown in Fig.~\ref{fig:tdcs-para}. They did not change 
within the thickness of the lines when 
an enlarged spatial box of $\xi_{\rm max}=150$ was used. 
They are numerically converged at $|m|_{\rm max}=5$ 
(see the convergence checks below).
Near the maximum of the TDCS, the time-independent ECS results are significantly
larger than the present FE-DVR and TDCC
predictions.
The present TDCS results for
the parallel geometry are approximately three times smaller than those of Morales
{\it et al} \cite{Morales2009}, but twice as large as those of Colgan
{\it et al} \cite{Colgan2008}, i.e., they generally lie in between the previous predictions. 
This is indicated by the scaling factors shown in the legend.

The various sets of results disagree in the predicted
magnitude as well as in the shape of the
angular dependence.  A prominent
difference concerns the ``wing''
structures of non\-vanishing TDCS values near $\theta_2\simeq 0^\circ$
and $360^\circ$ for $\theta_1=0^\circ$ and $30^\circ$ observed in
Ref.~\cite{Colgan2008}. In addition to the dominant back-to-back escape mode,
there is a noticeable forward emission in the
prediction of Colgan {\it et al}. 
This counter-intuitive forward-to-forward escape mode for two electrons carrying the same kinetic energy and traveling 
in the same direction is not supported by either the
ECS or the present FE-DVR calculations.

\begin{figure}[thb]
\centering
\epsfig{file=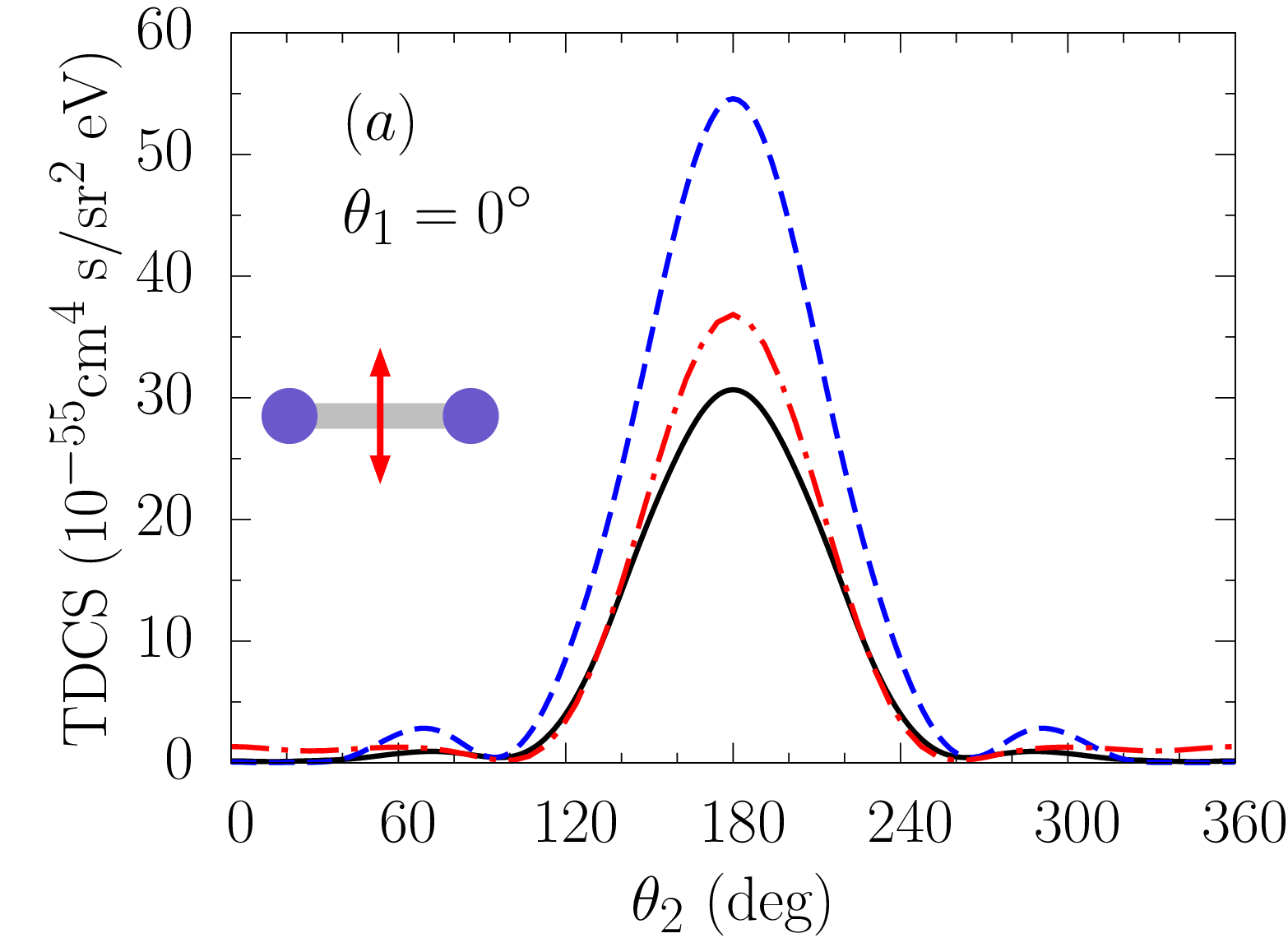,width=4.27cm,clip=}
\epsfig{file=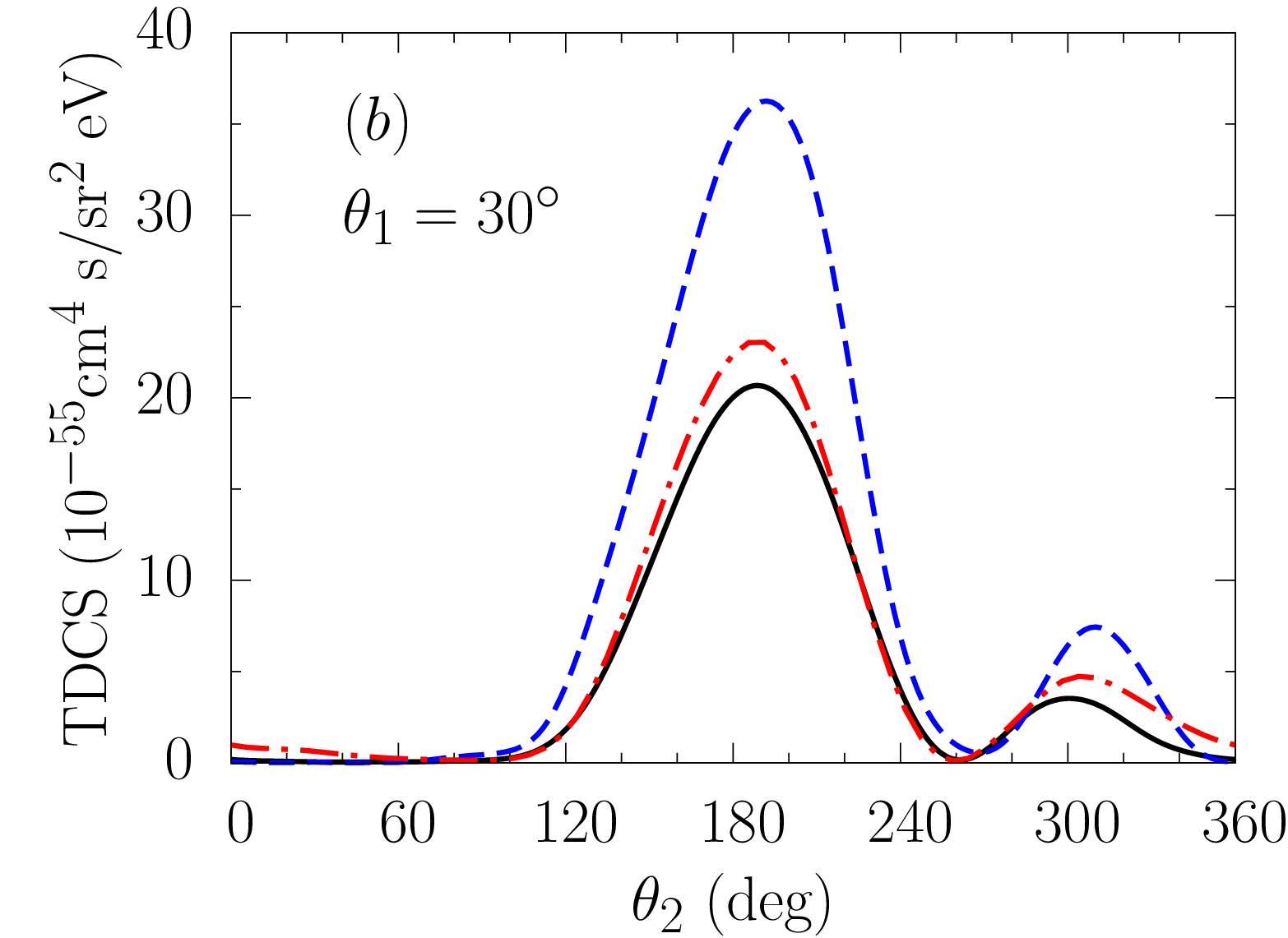,width=4.27cm,clip=} \\
\epsfig{file=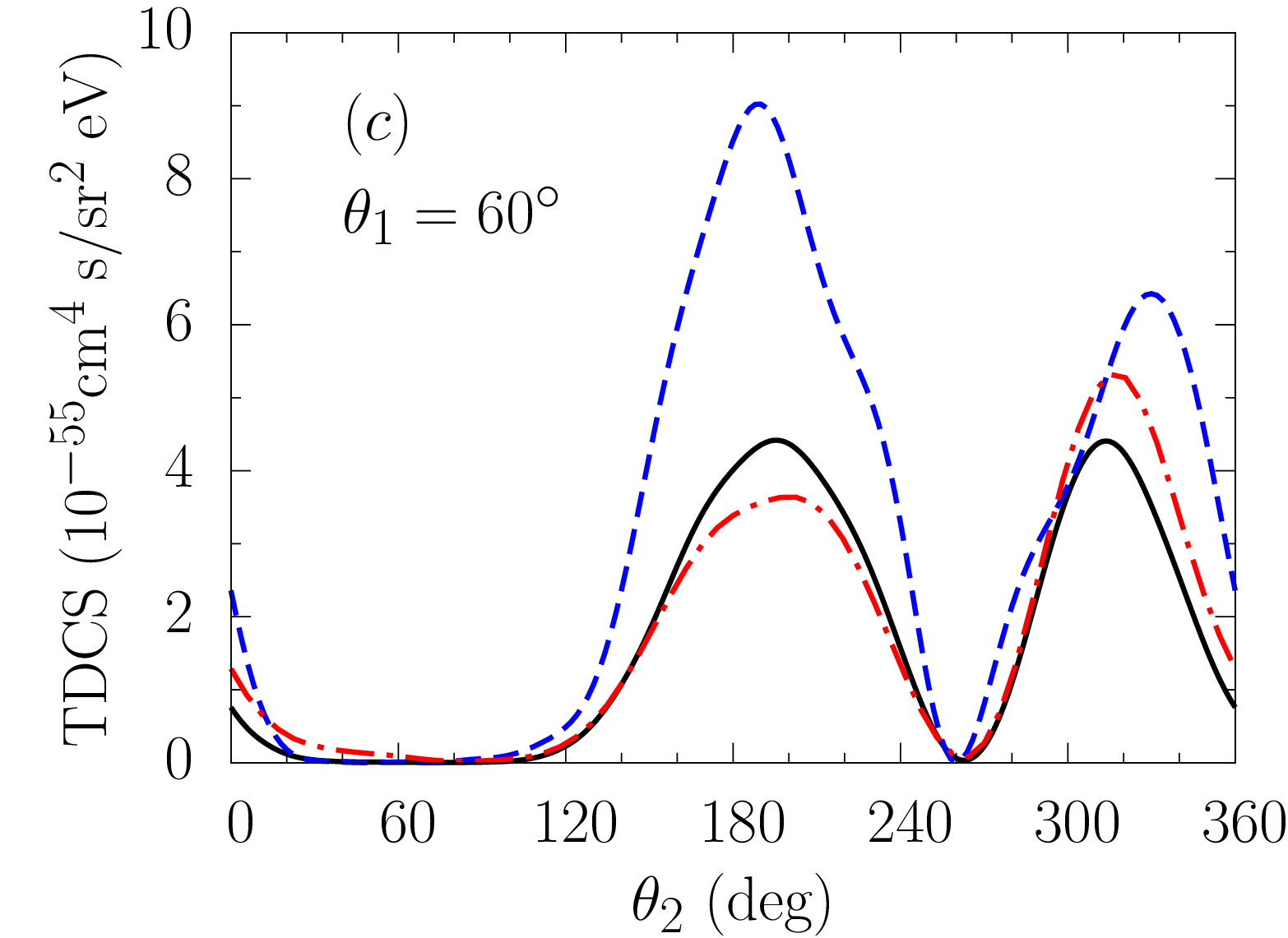,width=4.27cm,clip=}
\epsfig{file=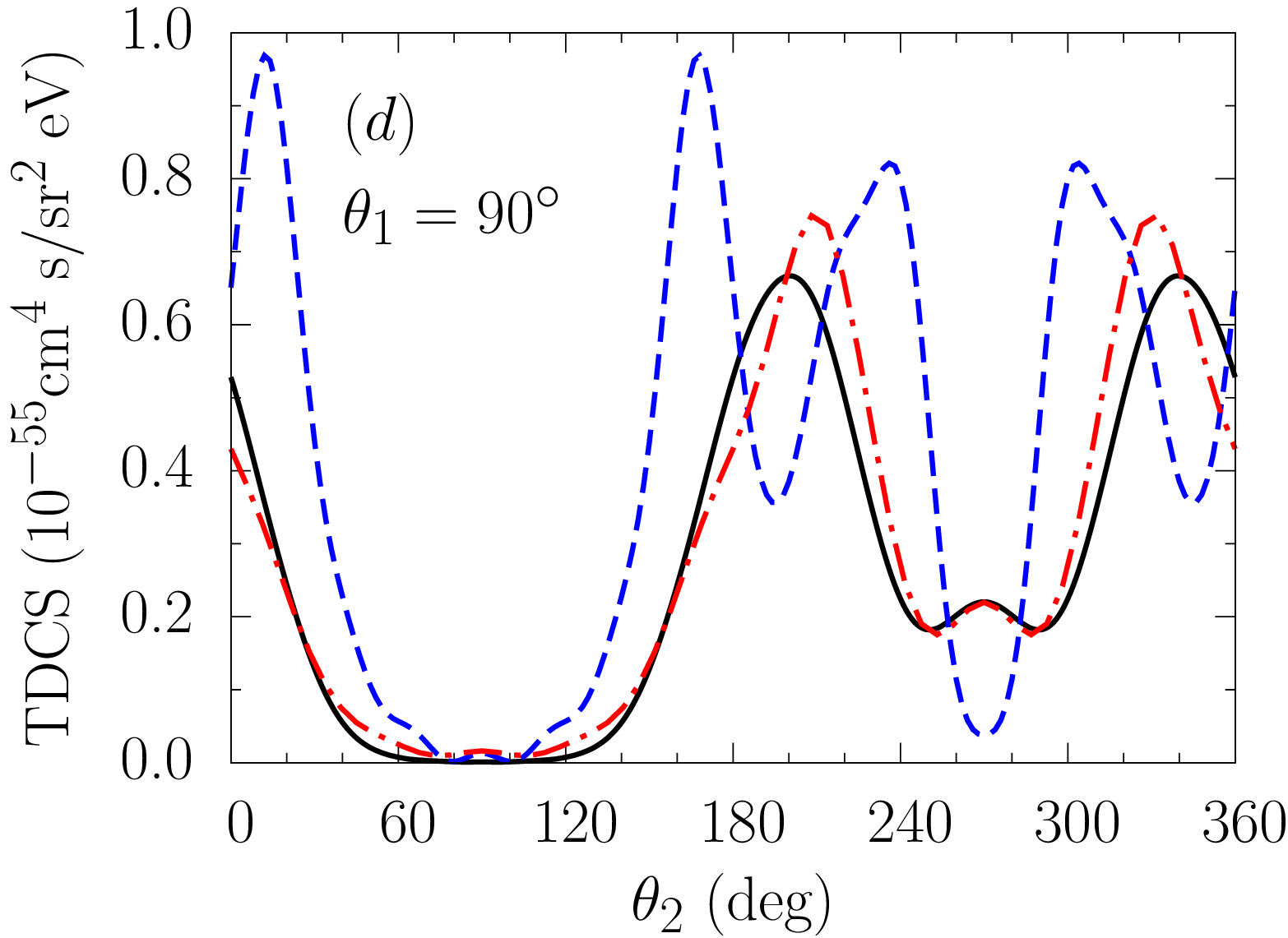,width=4.27cm,clip=}
\caption{(Color online) Same as Fig.~\ref{fig:tdcs-para} for the
perpendicular geometry.  No scaling factors were applied to compare
the various predictions.} 
\label{fig:tdcs-perp}
\end{figure}

Figure \ref{fig:tdcs-perp} displays the corresponding TDCSs for the perpendicular orientation.
Here the situation is different from the previous case. Although the ECS results~\cite{Morales2009} 
are generally larger than the present and the TDCC
predictions~\cite{Colgan2008}, they are of similar magnitude.  The
agreement between the FE-DVR and TDCC calculations is quite satisfactory in this case.
In contrast to the parallel geometry, the shapes of the angular
distributions in the perpendicular geometry are much closer to those obtained for the helium
atom at $42$~eV~\cite{Guan2008}.

The TDCS for orientation of the molecular axis parallel to the
polarization vector exhibits the molecular effect to the largest extent. 
Only the 
$\hbox{$^1\Sigma_g$}\rightarrow \hbox{$^1\Sigma_u$}\rightarrow \hbox{$^1\Sigma_g$}$ path of transitions is open in this geometry,
because the total magnetic quantum number along the ${\bm \varsigma}$-axis must be conserved. The final state, therefore, has the
same symmetry as the initial state.
At a right angle
between the ${\bm\epsilon}$- and ${\bm \varsigma}$-axes, on the other hand,
the two electrons may be
ionized through 
$\hbox{$^1\Sigma_g$}\rightarrow \hbox{$^1\Pi_u$}\rightarrow \hbox{$^1\Sigma_g$}$ or 
$\hbox{$^1\Sigma_g$}\rightarrow \hbox{$^1\Pi_u$}\rightarrow \hbox{$^1\Delta_g$}$, since the selection rule here is $|\Delta M|=1$
instead of $|\Delta M|=0$ (here $M=m_1+m_2$ is the total magnetic quantum number along the
$\bm{\varsigma}$-axis). This is qualitatively similar to those of the atomic target through 
$\hbox{$^1S^e$} \rightarrow \hbox{$^1P^o$}\rightarrow \hbox{$^1S^e$}$ or 
$\hbox{$^1S^e$} \rightarrow \hbox{$^1P^o$}\rightarrow \hbox{$^1D^e$}$. 
An interference effect between the
open channels $^1\Sigma_g$ and $^1\Delta_g$ in H$_2$ is evident, 
as it is in the helium atom between the $^1S^e$ and $^1D^e$ channels
\cite{Hu2005}. Such interference
cannot be observed
in the parallel orientation, where only one channel is open. 

\begin{figure}[t]
\centering
\epsfig{file=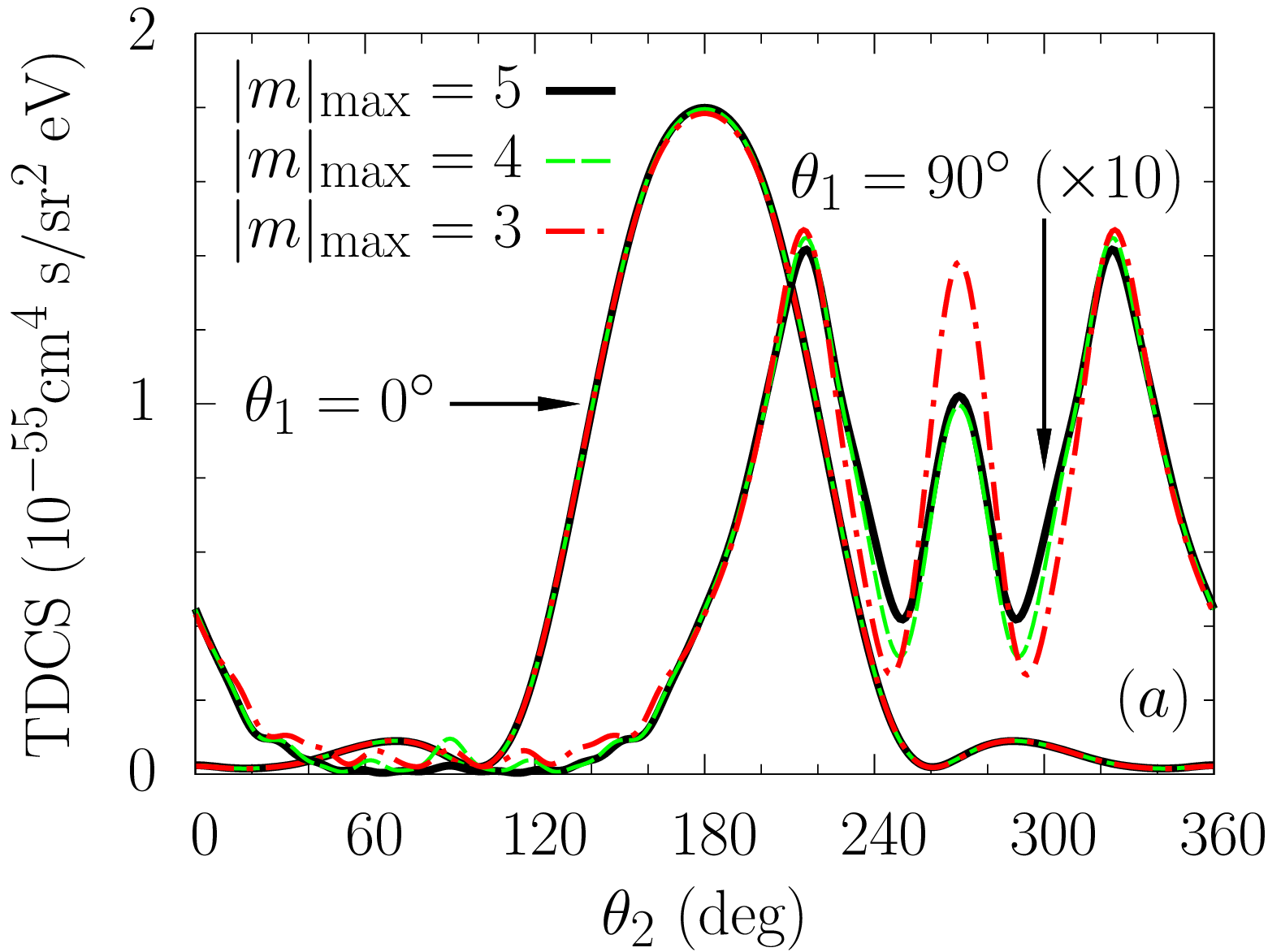,width=4.27cm,clip=} 
\epsfig{file=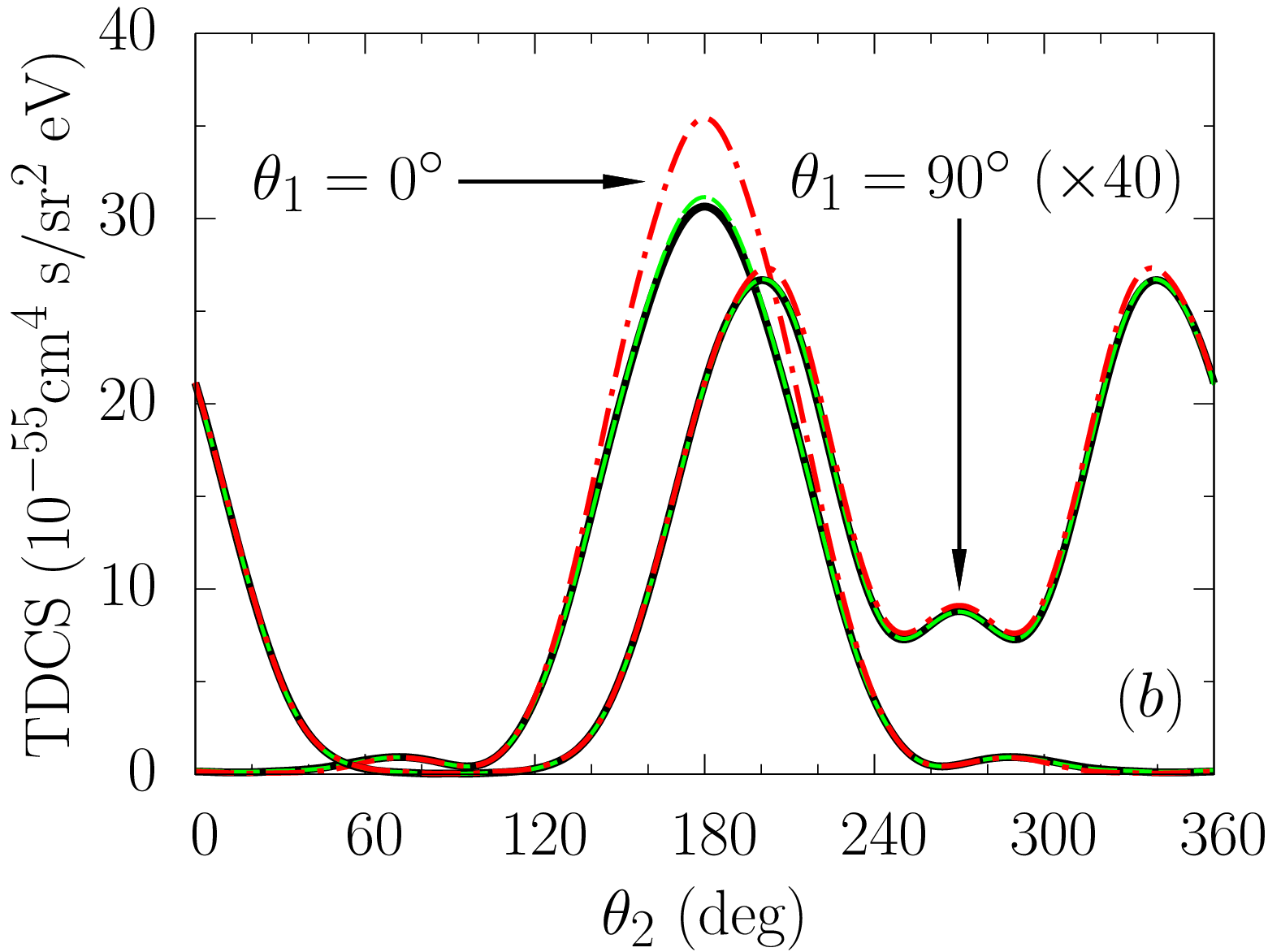,width=4.27cm,clip=}
\caption{(Color online) Convergence of the TDCS predictions with increasing
$|m|_{\rm max}$ for the parallel~(a) and perpendicular~(b)
geometry. The results for $|m|_{\rm max} = 4$ and~$5$ are hard to
distinguish.
The laser parameters are the same as in Fig.~\ref{fig:tdcs-para}.} 
\label{fig:tdcs-convg}
\end{figure}

We also performed extensive convergence checks
by increasing
$|m|_{\rm max}$ from $3$ to $5$.  Figure \ref{fig:tdcs-convg} shows the
convergence
for the fixed electron angles of $\theta_1 = 0^\circ$ and $\theta_1 = 90^\circ$. 
In the parallel case, the TDCS calculated with $|m|_{\rm max}=3$ at
$\theta_1=0^\circ$ is already fully converged, but for $\theta_1=90^\circ$ 
$|m|_{\rm max}= 5$ is required to obtain essentially converged results.
In the perpendicular case, on the hand, the TDCS at $\theta_1=0^\circ$ is very sensitive
to the value of $|m|_{\rm max}$, until it stabilizes at $|m|_{\rm max}=5$, while the TDCS at
$\theta_1=90^\circ$ is less sensitive to the choice of $|m|_{\rm max}$.

Although the parallel and perpendicular orientations exhibit different
convergence patterns
in terms of $\theta_1$ and $\theta_2$,
we observe that the TDCSs in both geometries behave similarly when we transfer the result to
the molecular frame.
In that frame, when the fixed electron is measured in the
direction perpendicular to the molecular axis, the back-to-back escape mode is most
sensitive to the $|m|$ value. When one electron is
detected along the molecular axis, the back-to-back mode is the most stable. 

In summary, we calculated the two-photon DI of the
H$_2$ molecule by an intense femto\-second laser pulse. 
Our TDCS results are generally closer to the predictions from the TDCC than from the ECS 
approach, except for the small-angle behavior in the parallel geometry, where we do not see what 
seems to be an unphysical increase of the TDCS.  A magnitude difference of about a factor of two, rather than five~\cite{Morales2009}, 
remains in the parallel case. 
Once again, however, we emphasize that a straight comparison of these results is inappropriate.
Great care should be taken when such a comparison is made, particularly when experimental data become available. 
Knowing the details of the experimental setup will be essential.
We plan to extend our present treatment to angular distributions in 
more general cases with an arbitrary alignment angle between the polarization
vector and the molecular axis, and to further study the sensitivity of the theoretical predictions on the details of the pulse.  

We are greatly indebted to Prof.~F.~Mart\'in for providing constructive remarks to
improve the manuscript.   We also thank Drs. J.~Colgan and F.~Morales for sending their
results in numerical form and helpful discussions. This work was supported by the NSF under grant 
PHY-0757755 (XG and KB) and supercomputer resources through the Teragrid allocation
TG-PHY090031.

\end{document}